\def\al{\alpha}
\def\be{\beta}

\def\de{\delta}
\def\ep{\epsilon}

\def\ka{\kappa}
\def\la{\lambda}

\def\si{\sigma}

\def\La{\Lambda}

\def\mn{{\mu\nu}}

\def\half{{\textstyle{1\over 2}}}
\def\frac#1#2{{\textstyle{{#1}\over {#2}}}}

\def\lsim{\mathrel{\rlap{\lower4pt\hbox{\hskip1pt$\sim$}}
    \raise1pt\hbox{$<$}}}
\def\gsim{\mathrel{\rlap{\lower4pt\hbox{\hskip1pt$\sim$}}
    \raise1pt\hbox{$>$}}}
\def\sqr#1#2{{\vcenter{\vbox{\hrule height.#2pt
         \hbox{\vrule width.#2pt height#1pt \kern#1pt
         \vrule width.#2pt}
         \hrule height.#2pt}}}}

\newcommand{\beq}{\begin{equation}}
\newcommand{\eeq}{\end{equation}}
\newcommand{\bea}{\begin{eqnarray}}
\newcommand{\eea}{\end{eqnarray}}

\documentclass{article}  

\begin{document}

\begin{center}
{{\bf Deformed Instantons \\}
\vglue 1.0cm
{Don Colladay and Patrick McDonald\\} 
\bigskip
{\it New College of Florida\\}
\medskip
{\it Sarasota, FL, 34243, U.S.A.\\}
 }
\end{center}
\vglue 0.8cm

\vglue 0.3cm
 
{\rightskip=3pc\leftskip=3pc\noindent
In this talk, instantons are discussed in the presence of Lorentz violation.  
Conventional topological arguments are applied to classify the
modified solutions to the Yang-Mills equations according to the 
topological charge.  Explicit perturbations to the instantons
are calculated in detail for the case of unit topological charge.
}

\vskip 1 cm

\vskip 1 cm

\section{Introduction}
Yang-Mills theories are typically constructed using a compact Lie Group
$G$ with Lie algebra $L(G)$, and a Lie algebra valued vector field $A^\mu(x)$.
The action of the group on the vector field is defined by
\beq
A^\mu (x) \rightarrow U(x) A^\mu (x) U^{-1}(x) - {i \over g} 
U(x) \partial^\mu U^{-1}(x) 
\quad ,
\eeq
where $U$ is a group element.
The field strength tensor is then defined as
\beq
F^\mn = - {i \over g} [D^\mu, D^\nu] 
\quad ,
\eeq
where the covariant derivative is taken as
$D^\mu = \partial^\mu + i g A^\mu$.
With this definition, the field strength transforms
in the simple way
\beq
F^\mn \rightarrow U(x) F^\mn U^{-1}(x)
\quad .
\eeq
The standard gauge invariant action is constructed 
by forming the integrated trace over the Lorentz invariant 
square of the field tensor
\beq
S_{YM}(A) = {1 \over 2} \int d^4 x ~ Tr[F_\mn F^\mn]
\quad .
\eeq
Extremization of this action with respect to $A$ yields
the equations of motion
\beq
[D_\mu, F^\mn] = 0
\quad .
\label{eom}
\eeq
The Bianchi identity follows from the definition
of the field strength and yields a further set of equations
\beq
[D_\mu, \tilde F^\mn] = 0
\quad ,
\label{bid}
\eeq
where $\tilde F^{\mn} = \half \ep^{\mn \al \be} F_{\al \be}$
is the dual of $F$.
Note that nonabelian groups yield nonlinear differential 
equations due to the nonvanishing field commutators.

It is possible that more fundamental theories of nature
may contain small Lorentz-breaking effects arising from
new physics at higher energy scales.\cite{kps}  
The Standard Model Extension (SME) provides a 
general framework within which to study Yang-Mills theory in the
presence of Lorentz violation.\cite{ck,cpt04}
This type of gauge theory has also been extended to 
include the gravitational sector.\cite{kgrav}
Including only gauge invariant and power-counting renormalizable 
corrections to the Yang-Mills sector yields the action
\bea
S(A) & = & {1 \over 2} \int Tr \left[ (F_\mn F^\mn) 
+ (k_F)^{\mn \alpha \beta} F_\mn F_{\al \be} \right.
\nonumber \\
& & \left. +  (k_{AF})^\ka  \ep_{\ka\la\mn}
(A^\la F^\mn - {2 \over 3} i g A^\la A^\mu A^\nu) \right]
\quad ,
\nonumber
\eea 
where $k_F$ and $k_{AF}$ are constant background fields that
parameterize the Lorentz violation.
The $k_{AF}$ terms present theoretical difficulties involving
negative energy issues\cite{cfj} even in the abelian case and are
therefore neglected in the following analysis.
The specific results discussed in this proceedings are derived
in more detail elsewhere.\cite{cm}

\section{Conventional Instantons}
In the standard case, static solutions to the Yang-Mills equations
with nontrivial, finite action (called instantons\cite{inst}) only occur 
when there are four spatial dimensions.\cite{deser}  Therefore, it is convenient
to transform the standard action to four-dimensional Euclidean space 
to perform the analysis:
\beq
S_0(A) = {1 \over 2} \int d^4 x ~ Tr[F^\mn F^\mn]
\quad ,
\eeq
where 
$F^\mn = \partial^\mu A^\nu - \partial^\nu A^\mu + i g [A^\mu, A^\nu]$ 
is the explicit form of the field tensor.
It is convenient to define a quantity
called the topological charge $q$ as
\beq
q = {g^2 \over 16 \pi^2} \int d^4 x Tr \tilde F^\mn F^\mn
\quad ,
\eeq
where 
$\tilde F^\mn = \frac 1 2 \ep^{\mn\al\be} F^{\al\be}$ 
is the dual of $F$.
Using the identity
$\frac 1 4 Tr \tilde F F = \partial^\mu X^\mu$
with
\beq
X^\mu = \frac 1 4 \ep^{\mn\la\ka}Tr(A^\nu F^{\la\ka} - \frac 2 3 i g A^\nu A^\la A^\ka)
\quad ,
\eeq
converts the integral to a surface integral.
The net result is that $q$ must be an integer that
represents the winding number of the group on 
the Euclidean three-sphere at infinity. 
Note that this argument is independent of the
explicit form of the action and only depends on the
asymptotic behavior of the fields.
The Euclidean version of the equations of motion (\ref{eom}) 
and the Bianchi Identity (\ref{bid}) yield a set of 
nonlinear coupled differential equations for $A^\mu$.
A clever argument for solving these equations \cite{bps}
has been developed.
The key identity in obtaining the instanton solutions is
\beq
\frac 1 2 \int d^4 x Tr(F \mp \tilde F)^2 \ge 0
\quad .
\eeq
This can be rearranged to yield the condition
\beq
S \ge \pm \frac 1 2 \int d^4 x Tr[\tilde F^\mn F^\mn] = \pm {8 \pi^2 \over g^2} q 
\quad .
\eeq
The inequality is saturated when the field strength satisfies
the duality condition $F = \pm \tilde F$.  
This means that if a self-dual (or anti-self-dual) field strength 
can be found, it will automatically extremize the action
and provide a solution to the equations of motion.

A theorem by Bott\cite{bott} states that any mapping of the 
Euclidean three-sphere into the group can
be continuously deformed into a mapping onto an SU(2)
subgroup.
It is therefore sufficient
to consider SU(2) subgroups of the full Lie group $G$.
An explicit example of a self-dual solution for $q=1$ is given by
\beq
A^\mu_{SD} = - {\tau^\mn x^\nu \over g(\rho^2 + x^2)}
\quad ,
\eeq
with associated field strength
\beq
F^\mn_{SD} = {2 \rho^2 \over g(\rho^2 + x^2)^2} \tau^\mn
\quad ,
\eeq
where $\tau^{0i} = \si^i$
and $\tau^{ij} = \ep^{ijk} \si^k$
are expressed in terms of the conventional Pauli matrices.
The anti-self-dual solution is the parity transform of 
the above solution.
Subsequent to this, all minimal action solutions have
been classified\cite{atiah} and formally constructed.

\section{Deformed Instantons}
When Lorentz violation is present, the Euclidean action is modified
as
\beq
S(A) = {1 \over 2} \int Tr[(F^\mn F^\mn) + (k_F)^{\mn \alpha \beta} 
F^\mn F^{\al \be}]
\quad .
\eeq 
Standard arguments demonstrate that instantons only exist in four
Euclidean dimensions, as in the conventional case.
The Lorentz violation is assumed small, therefore only
leading order contributions from $k_F$ are retained in the following
analysis.

As mentioned in the previous section, the topological charge
$q$ is an integer, regardless of the form of the action.  
A modified bound on the action can be derived as
\beq
S \ge \pm {8 \pi^2 \over g^2} q \pm \frac 1 4 
( k_F^{\mn\al\be} + \tilde k_F^{\mn \al \be}
)\int Tr \tilde F^\mn F^{\al\be}
\quad ,
\label{lbound}
\eeq
where $\tilde k_F^{\mn\al\be} 
= \frac 1 4 
\ep^{\mn\la\ka} k_F^{\la\ka\rho\si} 
\ep^{\rho\si\al\be}$
is a generalized dual to $k_F$.
This expression indicates the
natural decomposition
$k_F = k_{F}^+ \oplus k_{F}^-$ into its
self-dual and anti-self-dual parts.

\subsection{Case 1: $ k_F = - \tilde k_F $}

This condition implies that $k_F$
takes the form
$k_F^{\mn\al\be} = \La_k^{[\mu[\al} \de^{\nu]\be]} $
where $\La_k^\mn = \half k_F^{\al\mu\al\nu}$ depends
only on the trace components of $k_F$. 
The action can be minimized using the modified
duality condition
\beq
F^\prime \simeq \pm \tilde F^\prime
\quad ,
\eeq
where
$F^{\prime \mn} = F^\mn + \frac 1 2 k_F^{\mn\al\be} F^{\al\be}$.
The explicit solutions can be constructed using
the skewed coordinates
$\tilde x^\mu = x^\mu + \La_k^\mn x^\nu$
and writing
$A^\mu(x) \simeq A^\mu_{SD}(\tilde x) + \La^{\mn} A^\nu_{SD}(x)$
in terms of the modified coordinates.
These solutions therefore correspond to conventional instantons
in a skewed coordinate system.
Note that this is a result of the existence of field redefinitions
that can be used to transform physical effects of this type between 
the fermion and Yang-Mills sectors.\cite{cm2}

\subsection{Case 2: $ k_F = \tilde k_F $}
In this case, $k_F$ is trace free and there is no obvious 
modified duality condition on $F$ because
the lower bound on the action given in
Eq.(\ref{lbound}) varies with small fluctuations $\de F$,
therefore the equations of motion must be solved directly.
To find the deformation of the conventional instanton solution,
the potential is expanded as
$A = A_{SD} +  A_k$,
where $A_{SD}$ is the conventional self-dual solution and
$A_k$ is the unknown perturbation. 
The linearized equation of motion for $A_k$ is
\beq
[D_{SD}^\nu,[D_{SD}^\nu, A_k^\mu]] 
+ 2 i g [F_{SD}^\mn, A_k^\nu] 
= j_k^\mu
\quad ,
\eeq
where 
$j_k^\mu \equiv k_F^{\mn\al\be} 
[D^\nu_{SD}, F^{\al\be}_{SD}]$
is a set of known functions.
This gives a set of linear second-order
elliptic differential equations that can be
formally solved using propagator techniques:
\beq
A_k  = \int d^4 y G(x,y)  j_k(y)
\quad ,
\eeq
where $G$ is the appropriate Green's function.

As an explicit example, consider the deformation of a 
$q=1$ instanton in SU(2).
In this case, 
the direct Green's function approach is unwieldy,
therefore the following procedure was eventually adopted:\cite{cm}
\begin{itemize}
\item{Perform a gauge transformation to the singular 
gauge using $U(x) = -i x \cdot \tau^\dagger /x$ so that
the fields become quadratic in the instanton size.}
\item{Work to lowest order in the instanton size $\rho$
using the approximate Green's function 
$G^{-1} \simeq 4 \pi^2 (x - y)^2$, and integrate to find
the potential.}
\item{Use the tensorial structure of the resulting solution as
an ansatz for general values of $\rho$: 
\beq
A_k^\mu = {2 \rho^2 x^2 \over 3 g} f(x^2) 
k_F^{\mn\al\be}\tau^{\al(\nu} x^{\be)}
\eeq
Remarkably, this gives 
a differential equation for the unknown function $f(x^2)$.
}
\item{Solve the differential equation for $f$ to determine
the perturbation to all orders in $\rho$.} 
\end{itemize}
This solution explicitly preserves the topological charge
since the asymptotic fields at infinity and at the origin are unmodified. 
The structure of the instantons is only perturbed in the intermediate
region.

\section{Summary}
Instantons in the presence of Lorentz violation retain their topological
properties, but the detailed solutions are deformed.
Explicit solutions have been presented here for the case of
unit topological charge.
The deformations fall into two cases, one for which a simple redefinition
of coordinates provides solutions, and another that requires an explicit
solution to a set of linear second-order elliptic differential equations.
The explicit solution demonstrates that the instanton is unaltered
at the boundaries, but is deformed in the intermediate regions.


\begin{thebibliography}{xx}
\def\etal {{\it et al.}}

\bibitem{kps}
V.A.\ Kosteleck\'y and S.\ Samuel,
Phys.\ Rev.\ D {\bf 39}, 683 (1989);
{\it ibid.} 
{\bf 40}, 1886 (1989);
Phys.\ Rev.\ Lett.\ {\bf 63}, 224 (1989);
{\it ibid.} 
{\bf 66}, 1811 (1991);
V.A.\ Kosteleck\'y and R.\ Potting,
Nucl.\ Phys.\ B {\bf 359}, 545 (1991);
Phys.\ Lett.\ B {\bf 381}, 89 (1996);
Phys.\ Rev.\ D {\bf 63}, 046007 (2001); 
V.A.\ Kosteleck\'y, M.\ Perry, and R.\ Potting,
Phys.\ Rev.\ Lett.\ {\bf 84}, 4541 (2000). 

\bibitem{ck} 
D.\ Colladay and V.A.\ Kosteleck\'y,
Phys.\ Rev.\ D {\bf 55}, 6760 (1997);
Phys.\ Rev.\ D {\bf 58}, 116002 (1998).

\bibitem{cpt04}
For a summary of recent experimental tests
and theoretical progress, see, for example
V.A.\ Kosteleck\'y, ed.,
{\it CPT and Lorentz Symmetry II},
World Scientific, Singapore, 2002;
and this proceedings.

\bibitem{kgrav}
V.\ A.\ Kosteleck\'y,
Phys.\ Rev.\ D {\bf 69}, 105009 (2004).

\bibitem{cfj}
S.\ M.\ Carroll, G.\ B.\ Field, and R.\ Jackiw,
Phys.\ Rev.\ D {\bf 41}, 1231 (1990).

\bibitem{cm}
D.\ Colladay and P.\ McDonald,
J.\ Math.\ Phys.\ {\bf 45}, 3228 (2004).

\bibitem{inst}
For reviews, see, for example {\it Instantons in Gauge Theories}, ed. M. Shifman,
World Scientific, Singapore (1994); D. Freed and K. Uhlenbeck,
{\it Instantons and four-manifolds},
New York, Springer-Verlag (1991).

\bibitem{deser}
S.\ Deser,
Phys.\ Lett.\  {\bf 64B}, 463 (1976).

\bibitem{bps}
A.\ Belavin, A.\ Polyakov, A.\ Schwartz, 
and Y.\ Tyupkin,
Phys.\ Lett.\ {\bf 59B}, 85 (1975).

\bibitem{bott}
R.\ Bott,
Bull.\ Soc.\ Math.\ France {\bf 84}, 251 (1956).

\bibitem{atiah}
M.\ F.\ Atiyah, N.\ J.\ Hitchin,
V.\ G.\ Drinfeld, and Y.\ I.\ Manin,
Phys.\ Lett.\ {\bf 65A} 285 (1978).

\bibitem{cm2}
D.\ Colladay and P.\ McDonald,
J.\ Math.\ Phys.\ {\bf 43}, 3554 (2002); 
V.\ A.\ Kosteleck\'y and M.\ Mewes,
Phys.\ Rev.\ D {\bf 66}, 056005 (2002).

\end{thebibliography}
\end{document}